\documentclass[aps,prl,twocolumn,groupedaddress,showpacs]{revtex4-1}

\usepackage{graphicx}
\usepackage{color}

\begin{document}


\title{A Big-Bang Nucleosynthesis Limit on the Neutral Fermion Decays into Neutrinos}


\author{Motohiko Kusakabe$^{1,2}$}
\email{motohiko@kau.ac.kr} 
\author{A.B. Balantekin$^{3,4}$}
\email{baha@physics.wisc.edu}
\author{Toshitaka Kajino$^{4,5}$}
\email{kajino@nao.ac.jp}
\author{Y. Pehlivan$^{4,6}$}
\email{ypehlivan@me.com}
\affiliation{
$^1$School of Liberal Arts and Science, Korea Aerospace University, Goyang 412-791, Korea}
\affiliation{
$^2$Department of Physics, Soongsil University, Seoul 156-743, Korea}
\affiliation{
$^3$Department of Physics, University of Wisconsin, Madison, WI 53706, USA}
\affiliation{
$^4$National Astronomical Observatory of Japan 2-21-1 
Osawa, Mitaka, Tokyo, 181-8588, Japan}
\affiliation{
$^5$Department of Astronomy, Graduate School of Science, 
University of Tokyo, 7-3-1 Hongo, Bunkyo-ku, Tokyo, 113-0033, Japan}
\affiliation{
$^6$Mimar Sinan Fine Arts University, Sisli, Istanbul 34380, Turkey
}



\date{\today}

\begin{abstract}
Using the primordial helium abundance, an upper limit to the magnetic moments for Dirac neutrinos had been provided by imposing restrictions on the number of the additional helicity states. Considering non-thermal photons produced in the decay of the heavy sterile mass eigenstates due to the neutrino magnetic moment, we explore the constraints imposed by the observed abundances of all the light elements produced during the Big Bang nucleosynthesis. 
\end{abstract}


\maketitle


Neutrino magnetic moments are expected to be very small. Although one can enumerate various terrestrial experiments that can probe the neutrino magnetic moment \cite{Vogel:1989iv}, the best limits on this quantity comes from the scattering of neutrinos or antineutrinos off electrons. At low enough electron recoil energies, the magnetic moment contribution to the cross section exceeds the standard weak contribution with the lowest measurable recoil energy providing the best limit. The best direct limits on the neutrino magnetic moment come from experiments with reactor antineutrinos \cite{Beda:2012zz,Deniz:2009mu}. The current limit is $\mu_{\nu} < 2.9 \times 10^{-11} \mu_B$ at 90\% C.L. where $\mu_B = e/2m_e$ is the Bohr magneton. 

Further constraints on the neutrino magnetic moment come from astrophysical and cosmological arguments. The most strongest such limit comes from the cooling of red giant stars. Before the helium flash, the degenerate helium core loses energy by neutrino pair emission. If there is a sizable neutrino magnetic moment, besides the Standard Model processes, additional cooling of the core is possible through plasmon decay into neutrino pairs. Since the cross section of the latter process is proportional to the $\mu_{\nu}^2$, a large magnetic moment will delay the helium ignition, altering the ratio of red giant to horizontal branch stars. Observations of globular-cluster stars  result in a limit of  $\mu_{\nu} < 3 \times 10^{-12} \mu_B$ \cite{Raffelt:1990pj}.  Another, less stringent, argument comes from the observation of the neutrinos from SN1987A \cite{Lattimer:1988mf}. Magnetic moment contribution to the neutrino scattering is mediated by a photon exchange, hence it changes the helicity of the neutrino. If the neutrinos are of Dirac type, the right-handed states are sterile and can easily escape the core. If the magnetic moment is relatively large this mechanism would dominate the cooling rate. The low energies of the neutrinos and the multi second time-scale of the burst observed in SN1987A is in accordance with a diffusely cooling protoneutron  star, suggesting that no such right-handed states are created. However, such a scenario assumes that the dynamics of the core-collapse supernovae is well understood. This limit does not apply to Majorana neutrinos since their right-handed counterparts are not sterile. 

Cosmological arguments given so far are not very constraining either. In the standard description of the early universe, during the nucleosynthesis epoch neutrinos are assumed to have decoupled when $T \sim 2 m_e$ and subsequently only electron-positron pairs 
interact with the photons. Clearly a large enough magnetic moment could permit neutrino-photon interactions at later times. However, it turns out that primordial helium synthesis is more sensitive to the additional neutrino helicities than considerations of neutrino equilibration. Imposing the requirement that the helium synthesis in the Big Bang not to be disrupted by the production of additional right-handed states limits the neutrino magnetic moment to be $\mu_{\nu} < 10^{-11} \mu_B$ \cite{morgan}. A more careful treatment of the conditions in the Big Bang may loosen this limit by about a factor of three  \cite{Fukugita:1987uy}. Again this limit does not apply to Majorana neutrinos since no additional states are produced with their magnetic interactions. It should be emphasized that both the globular cluster and supernova limits are restricted to neutrinos (active or sterile) light enough to be produced in these environments (less than a few keV for the helium flash and less than a few MeV for the supernovae). 

Anomalous results from a variety of neutrino experiments could be interpreted as evidence for the existence of sterile neutrinos, additional neutrino mass states beyond the three active species in the Standard Model. Recently, a reanalysis of short-baseline reactor neutrino experiments has revealed a discrepancy between observations and the expected antineutrino flux \cite{Mention:2011rk}. While a full  resolution of this discrepancy requires further experimental work \cite{Heeger:2012tc}, it has renewed interest in light sterile neutrinos (for a recent review see Ref. \cite{Abazajian:2012ys}). It is not easy to see if such light sterile neutrino states can be accommodated in standard cosmology \cite{Hamann:2011ge}. It is prudent to explore the implications of the heavier (with masses more than a few MeV) sterile neutrinos to ensure that our theoretical understanding is as complete as possible and there are no additional effects that may impact active-sterile mixing. To this end we explore the implications of heavier sterile states for the Big Bang nucleosynthesis  (BBN). 

If neutrinos possess non-zero magnetic moments,  a sterile state would decay into another sterile or active state by photon emission. The photon emitted in this decay during the  BBN epoch is not likely to be in thermal equilibrium. 
The radiative lifetime of  such a sterile neutrino is given as (see e.g. \cite{Raffelt:1999gv})
\begin{eqnarray}
\label{1}
\tau_X^{-1} &=& \frac{|\mu_{ij}|^2 + |\epsilon_{ij}|^2}{8\pi} \left( \frac{m_i^2 - m_j^2}{m_i} \right)^3
\nonumber \\
&=& 5.308 s^{-1} \left( \frac{ \mu_{\rm eff}}{\mu_B} \right)^2 \left( \frac{m_i^2 - m_j^2}{m_i^2} \right)^3 \left( \frac{m_i}{{\rm eV}}\right)^3. 
\end{eqnarray}
In this expression, $\mu_{ij}$ and $\epsilon_{ij}$ are the magnetic and electric dipole moments that take us from the heavy mass eigenstate $i$ to the light mass eigenstate $j$. In this paper we use the shorthand notation $|\mu_{\rm eff}|^2 = |\mu_{ij}|^2 + |\epsilon_{ij}|^2$. Eq. (\ref{1}) can be derived from an effective Lagrangian that describes the electromagnetic coupling between a heavy neutral fermion ("sterile neutrino") and a light (active) mass eigenstate. 
In the discussion below, we write the mass of the decaying  state generically as $m_X$.  We will assume that there is only one sterile state.  

Reactor experiments probing the neutrino magnetic moment are inclusive experiments, and they do not observe the outgoing neutrino. Hence, if one neglects all the neutrino masses, they measure the quantity \cite{Balantekin:2006sw}
\begin{equation}
\label{2}
(\mu^2)_e \sim \sum_{ijk} U_{ei} \mu_{ik} \mu^{\dagger}_{kj} U^{\dagger}_{je} = (U\mu \mu^{\dagger}U^{\dagger})_{ee}. 
\end{equation}
If the neutrino masses are sizable there are small corrections to Eq. (\ref{2}). Note that this quantity is not simply the square of the diagonal magnetic moment of the electron. The magnetic moment can change flavor and all flavors permitted by the kinematics can be created and summed over in the final state. Indeed for the Majorana neutrinos only non-diagonal magnetic moments exist. 
Note that a sterile neutrino that mixes with active flavors contains all the mass eigenstates:
\begin{equation}
|\nu_{\rm sterile} \rangle = \sum_i U_{si} | \nu_i \rangle .
\end{equation}

If we ignore the mass of final mass eigenstate ($m_j \sim 10^{-2}$ eV) in Eq. (\ref{1}), then the energy of the produced photon is
\begin{equation}
E_{\gamma 0} = \frac{1}{2} \sqrt{p^2 + m_i^2} ,
\end{equation}
where $p$ is the initial momentum of the neutrino.  As we mentioned above, such photons will not be in thermal equilibrium. It is known that 
such nonthermal photons can induce electromagnetic cascade showers, and generate many less energetic nonthermal photons (e.g.,~\cite{Ellis:1984er,Kawasaki:1994sc}). If the sterile species decay after the $e^+e^-$ annihilation occurs, these nonthermal photons can disintegrate background light elements, potentially altering abundances of nuclei synthesized during the Big Bang~\cite{Lindley1979MNRAS.188P..15L,Ellis:1984er,Dimopoulos:1987fz,1992NuPhB.373..399E,Kawasaki:1994af,Kawasaki:1994sc,Jedamzik:1999di,Kawasaki:2000qr,Cyburt:2002uv,Kawasaki:2004qu,Ellis:2005ii,Jedamzik:2006xz,Kusakabe:2006hc}.  

In our calculations we utilized Kawano's public BBN code~\cite{Kawano1992,Smith:1992yy}, and adopted Sarkar's correction for $^4$He abundances~\cite{Sarkar:1995dd}.  Reaction rates for light nuclei ($A \le  10$) are updated with recommendations by JINA REACLIB Database V1.0 \cite{Cyburt2010}, and the neutron lifetime of $878.5 \pm 0.7_{\rm stat} \pm 0.3_{\rm sys}$~s~\cite{Serebrov:2010sg,Mathews:2004kc} based on improved measurements \cite{Serebrov:2004zf} was adopted.  We adopt the method of  Ref.~\cite{Kusakabe:2006hc} to calculate the nonthermal nucleosynthesis, where thermonuclear reactions are simultaneously taken into account.  We took updated reaction rates of $^4$He photodisintegration [eqs. (2) and (3) of Ref. \cite{Kusakabe:2008kf}] using the cross section data from precise measurements with laser-Compton photons~\cite{Shima:2005ix,Kii:2005rk}.  

In earlier studies of the nonthermal nucleosynthesis, the mass of the decaying particle is usually assumed to be larger than $\mathcal{O}$(1 MeV), which is the scale of threshold energies of nuclear photodisintegration.  However, in this study, this restriction is not imposed in order to analyze the mass-dependent effects of the radiative decay  for the first time.  In this case, there are three parameters.  The first parameter is $(n_X^0/n_\gamma^0)$, the number ratio of the decaying sterile neutrino state $\nu_s$ to the background  radiation before the decay of $\nu_s$.  The second parameter is $\tau_X$, the lifetime of the decaying eigenstate, or equivalently the neutrino magnetic moment as given in Eq. (\ref{1}).  The third parameter is $E_{\gamma 0}$, the energy of photon emitted at the radiative decay.

The relic abundance of extra neutrino $X$ is determined as follows.
The decoupling of neutrino from the thermal bath in the early universe occurs at temperature $T_\mathrm{d}\sim 1$ MeV for light sterile species ($m_X \ll 1$ MeV), and at $T_\mathrm{d}\simeq m_X/20$ for heavy sterile species $(m_X \gg 1$ MeV) \cite{Ressell:1989rz}.  It is then simply assumed that massive neutrinos decouple at the higher of these two temperatures, i.e., $T_{\rm d}=$max($1$ MeV, $m_X/20$).  The neutrino abundance at the decoupling time is then given by
\begin{equation}
\label{5} 
n_{\mathrm{d}X}(m_X)=\frac{g_X}{2\pi^2} \int_0^\infty dp \frac{p^2}{\exp\left[\sqrt[]{\mathstrut p^2+m_X^2}/T_\mathrm{d}(m_X)\right]+1},
\end{equation}
where
$g_X$ is the spin degrees of freedom of the species $X$, which is assumed to be unity here.
In the epoch of $e^+~e^-$ annihilation, the number density of neutrinos is diluted by the annihilation.  The number ratio of neutrinos and photons at $T\ll 1$ MeV is given by
\begin{equation}
\frac{n_X}{n_\gamma}=\frac{4}{11}\frac{n_{\mathrm{d}X}(m_X)}{n_\gamma(T_\mathrm{d})}=\frac{2\pi^2}{11\zeta(3)} \frac{n_{\mathrm{d}X}(m_X)}{T_\mathrm{d}^3}.
\end{equation}

The photon emitted at the radiative decay loses its energy through interactions with background photons, and electromagnetic showers composed of energetic photons, electrons and positrons are induced.  When the energy of the emitted photon $E_{\gamma0}$ is much larger than the threshold energy of photodisintegration of light nuclides, i.e., $E_{\gamma0} \gg 1$ MeV, the steady state energy spectrum of the nonthermal photons is approximately given (e.g., \cite{Cyburt:2002uv,Kusakabe:2006hc}) by
\begin{equation}
p_\gamma(E_\gamma)=\left\{
\begin{array}{ll}
K(E_X/E_\gamma)^{3/2} & \mathrm{for}~E_\gamma < E_X,\\
K(E_X/E_\gamma)^2 & \mathrm{for}~E_X < E_\gamma < E_C,\\
0 & \mathrm{for}~E_C < E_\gamma,\\
\end{array}
\right.
\label{eq_spe}
\end{equation}
where 
$E_X\sim m_e^2/(80T)$ and $E_C\sim m_e^2/(22T)$ are the energy corresponding to a break in the power law, and a cutoff energy, respectively,~\cite{Kawasaki:1994sc} with $m_e$ the electron mass,
$K=E_{\gamma0}/\{E_X^2[2+\ln(E_C/E_X)]\}$ is the normalization constant which conserves the energy of
the initially injected photons. Note that the spectrum has a cutoff at the energy $E_C$ because for energies larger than $E_C$, photons
are quickly destroyed in electron-positron pair production. 
The spectrum given in Eq. (\ref{eq_spe}) is valid when $E_{\gamma0} \gg 1$ MeV but in this study we are also
concerned with situations where the emitted photon energy is near the photodisintegration threshold, i.e., 
$E_{\gamma0}={\mathcal O}$(1 MeV).  To accommodate these cases, we implement the following generalization:  
\\
(1)  if $E_{\gamma0}<E_X$, then the spectrum is given by 
\begin{equation}
p_{1 \gamma}(E_\gamma)=\left\{
\begin{array}{ll}
K_1(E_X/E_\gamma)^{3/2} & \mathrm{for}~E_\gamma < E_{\gamma0},\\
0 & \mathrm{for}~E_{\gamma0} < E_\gamma,\\
\end{array}
\right.
\end{equation}
where
$K_1=E_{\gamma0}^{1/2}/(2 E_X^{3/2})$,
\\
(2)  if $E_X< E_{\gamma0}<E_C$, then the spectrum is given by
\begin{equation}
p_{2 \gamma}(E_\gamma)=\left\{
\begin{array}{ll}
K_2(E_X/E_\gamma)^{3/2} & \mathrm{for}~E_\gamma < E_X,\\
K_2(E_X/E_\gamma)^2 & \mathrm{for}~E_X < E_\gamma < E_{\gamma0},\\
0 & \mathrm{for}~E_{\gamma0} <E_\gamma,\\
\end{array}
\right.
\end{equation}
where
$K_2=E_{\gamma0}/\{E_X^2[2+\ln(E_{\gamma0}/E_X)]\}$, and 
\\
(3)  if $E_C< E_{\gamma0}$, then the spectrum $p_{3 \gamma}(E_\gamma)$ is the
one given in Eq. (\ref{eq_spe}).

Cosmological parameters are taken from the analysis of the Wilkinson Microwave Anisotropy Probe (WMAP)
\cite{Spergel:2003cb,Spergel:2006hy,Larson:2010gs,Hinshaw:2012fq}.  We adopt central values of constrained parameter
regions for model $\Lambda$CDM (WMAP only) determined from WMAP9 data:  $H_0=70.0 \pm 2.2$ km/s/Mpc, $\Omega_{\rm
b}=0.0463 \pm 0.0024$, $\Omega_\Lambda=0.721 \pm 0.025$, $\Omega_{\rm m}=0.279 \pm 0.025$, and $\eta=(6.19\pm 0.14)
\times 10^{-10}$~\cite{Hinshaw:2012fq}.

 By taking into account the results of a recent detailed calculation for the
conversion of $^7$Be to $^7$Li in the recombination epoch (Fig. 4 of Ref. \cite{Khatri:2010ed}), we assume
an approximately instantaneous conversion via the electron capture of $^7$Be$^{4+}$ at the redshift $z=3\times 10^4$.

Calculated results are compared with observational constraints on elemental
abundances.  The primordial D abundance is inferred from observations of QSO
absorption systems including a damped Lyman alpha system of QSO SDSS J1419+0829, which is
measured most precisely~\cite{Pettini:2012ph}.  We adopt the mean value estimated from ten
Lyman-$\alpha$ absorption systems, log(D/H)$=-4.58\pm 0.02$~\cite{Pettini:2012ph}.
Allowing for $2\sigma$ uncertainty, we take the constraint as
$2.40\times10^{-5}<$D/H$<2.88\times10^{-5}$.  $^3$He abundances in Galactic HII regions
have been measured through the $8.665$~GHz hyperfine transition of $^3$He$^+$,
$^3$He/H$=(1.9\pm 0.6)\times 10^{-5}$~\cite{Bania:2002yj}.  We take the $2\sigma$ upper
limit and adopt $^3{\rm He}/{\rm H}< 3.1\times 10^{-5}$.

The primordial $^4$He abundance is inferred from observations of metal-poor extragalactic
HII regions.  Two different determinations have been published recently, $Y_{\rm
p}=0.2565\pm 0.0051$~\cite{Izotov:2010ca} and $Y_{\rm p}=0.2561\pm
0.0108$~\cite{Aver:2010wq}.  We take the latter conservative constraint and allow for $2\sigma$
uncertainty:  $0.2345 < Y_{\rm p} < 0.2777$~\cite{Aver:2010wq}. 

Primordial $^7$Li abundance is inferred from spectroscopic observations of metal-poor
stars.  The observed abundances are different from theoretically calculated
values in the standard BBN model.  Although the cause of this disagreement
is not clear yet \cite{Erken:2011vv,Kusakabe:2012ds}, we adopt the observational abundance, log($^7$Li/H)$=-12+(2.199\pm
0.086)$ derived in a 3D nonlocal thermal equilibrium model~\cite{Sbordone2010}. 
Allowing for $2\sigma$ uncertainty, the constraint is $1.06\times
10^{-10}<^7$Li/H$<2.35\times 10^{-10}$.

In some of metal-poor stars, $^6$Li is detected.  The most probable detection in G020-024
shows the isotropic ratio of $^6$Li/$^7$Li=$0.052\pm 0.017$ \cite{2010IAUS..265...23S}.
We use the $2\sigma$ upper limit and $\log(^7{\rm Li}/{\rm H})=-12+2.18$ for the same star
\cite{Asplund:2005yt}, and adopt the constraint, $^6$Li/H$<1.3\times 10^{-11}$.

The radiative decay of massive neutrino enhances the energy of cosmic background radiation or entropy density.  The baryon-to-photon number ratio then changes as a function of cosmic time \cite{Feng:2003uy}.   When nonthermal photons are injected following the sterile neutrino decay, which occurs before the cosmological recombination, the comoving entropy of the universe increases.  The baryon-to-photon ratio is inversely proportional to the comoving entropy so that it is reduced during the nonthermal photon injection.  However, since the baryon-to-photon ratio inferred from WMAP  measurement of cosmic microwave background anisotropies is consistent with primordial elemental abundances, the entropy change is constrained.  When the change in comoving entropy ($S$) is small, the ratio of comoving entropies measured before and after the radiative decay, i.e., $S_{\rm i}$ and $S_{\rm f}$, is approximately given \cite{Feng:2003uy} by 

\begin{equation}
\frac{S_{\rm f}}{S_{\rm i}}=\exp\left[\frac{(45)^{3/4} \zeta(3)}{\pi^{11/4}}~\frac{\left(g_\ast^{\tau_X}\right)}{g^i_{\ast S}} 
~\frac{E_{\gamma 0}  n_X^{\rm  i}}{n_\gamma^{\rm i}} ~\sqrt{\frac{\tau_X}{M_{\rm Pl}}}
\right],
\end{equation}
where 
$\zeta(3)=1.202$ is the zeta function; 
$g_\ast^{\tau_X}=3.36$ and $g_{\ast S}^{\tau_X}=3.91$ are relativistic degrees of freedom in terms of number and entropy, respectively; 
$n_{\rm X}^{\rm i}$ and $n_\gamma^{\rm i}$ are number densities of the decaying sterile species and photon, respectively, evaluated at the same time before the decay; 
and $M_{\rm Pl}=1.22\times 10^{19}$ GeV is the Planck mass.  In the limit of small fractional change of entropy, the value is given \cite{Feng:2003uy} by
\begin{equation}
\frac{\Delta S}{S}\approx \ln \frac{S_{\rm f}}{S_{\rm i}}=1.1\times 10^{-4}
\left(\frac{{E_{\gamma 0} n_X^{\rm i}/n_\gamma^{\rm i}}}{10^{-9}}\right)^{1/4} \left(\frac{\tau_X}{10^6~{\rm s}}\right)^{1/2},
\end{equation}
Using the $2\sigma$ uncertainty from the baryon number density measured by WMAP, we allow at most $10 \%$ change in comoving entropy.

\begin{figure}
\begin{center}
\includegraphics[width=6.0cm,clip]{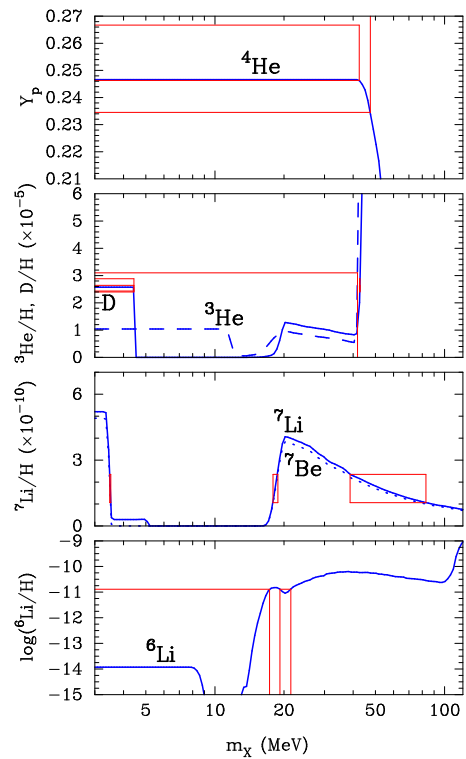}
\caption{Calculated nuclear abundances in the radiatively decaying heavy neutrino model as functions of the neutrino mass for a fixed lifetime of $\tau_X=10^{10}$ s.  Solid lines show final abundances of $^4$He, D, $^7$Li, and $^6$Li, while the dashed line shows final abundance of $^3$He.  The dotted line corresponds to the $^7$Be abundance before it is converted to $^7$Li via the electron capture process.  Boxes indicate observational constraints.  \label{fig1}}
\end{center}
\end{figure}

Figure \ref{fig1} shows calculated abundances in the radiatively decaying neutrino model as a function of mass $m_X$ for a fixed lifetime of $\tau_X=10^{10}$ s.  This lifetime is chosen merely to illustrate the way constraints are obtained.  Final abundances of $^4$He, D, $^7$Li, and $^6$Li, (solid lines) and $^3$He (dashed line) are shown.  The dotted line is for $^7$Be abundance before it is converted to $^7$Li via electron capture process operating in the recombination epoch of $^7$Be$^{4+}$~\cite{Khatri:2010ed}.

In drawing Fig. \ref{fig1}, we calculated transfer functions of nonthermal nuclei which depend on the mass of the decaying particle $X$.  
The following analysis of the time evolution of abundances for different masses is useful in understanding Fig. \ref{fig1}:
The $^4$He abundance decreases as a function of $m_X$ in the region of $m_X \geq 40$ MeV by photodisintegration reactions.  The D abundance decreases at $m_X \geq 4.4$ MeV by photodisintegration, and increases at $m_X \geq 40$ MeV because of a production via the photodisintegration of $^4$He.  The $^3$He abundance decreases at $m_X \geq 11$ MeV by photodisintegration, and increases at $m_X \geq 40$ MeV from the photodisintegration of $^4$He.  The curves for $^7$Li (the sum of abundances of $^7$Li and $^7$Be) and $^7$Be reflect the photodisintegration of $^7$Li ($m_X \geq 5$ MeV) and $^7$Be ($m_X \geq 3.2$ MeV).  The $^6$Li abundance decreases at $m_X \geq 9$ MeV by photodisintegration, and increases by production from the $^7$Be photodisintegration ($m_X \gtrsim 20$ MeV) and the $^4$He photodisintegration followed by $\alpha$-fusion reactions, $^4$He($\gamma$,$p$)$^3$H($\alpha$,$n$)$^6$Li, and $^4$He($\gamma$,$n$)$^3$He($\alpha$,$p$)$^6$Li ($m_X \gtrsim 100$ MeV).

\begin{figure}
\begin{center}
\includegraphics[width=8.0cm,clip]{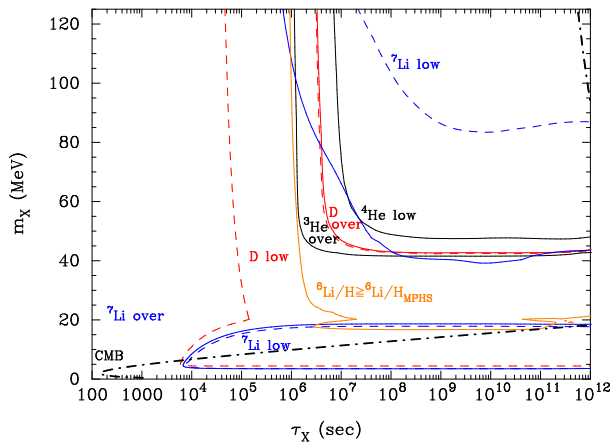}
\caption{Contours in the ($\tau_X$, $m_X$) plane for the adopted constraints of the primordial abundances of D (red), $^3$He and $^4$He (black), and $^7$Li (blue).  The right regions of solid lines for D, $^3$He, $^4$He, and dashed line for $^7$Li are excluded because of too large effects on nuclear abundances.  The region bounded by two dashed lines for D is excluded due to underproduction of D. $^7$Li is overproduced on the left of the solid $^7$Li line. Regions with the notation, over and low, indicate that they are excluded by overproduction and underproduction, respectively.  The $^6$Li abundance is larger than that detected in the metal-poor star in the right region from the curve for $^6$Li.  The right region from the dot-dashed line is excluded by the change of baryon-to-photon ratio. \label{fig2}}
\end{center}
\end{figure}

Apart from the discrepancy between theoretical and observational $^7$Li abundances, the constraint on the mass, $m_\nu<4.4$ MeV, is derived for $\tau_X=10^{10}$ s.  This bound comes from the constraint on D abundance.  Interestingly, there is a solution to the $^7$Li abundance problem at $m_\nu\sim$ 3.5 MeV.  In this region, the photodisintegration of $^7$Be is induced while the photodisintegration of other light nuclei never occurs since the energy of nonthermal photons is always below the reaction thresholds.  The possibility of this finely tuned mass of the decaying particle is noted as a "just-so" solution in Ref. \cite{Pospelov:2010hj}, although a quantitative calculation has never been reported.  In fact, this parameter region is excluded in Fig. \ref{fig2} when the entropy production associated with $X$ decay is considered.  However, if the relic abundance of the $X$ particle is smaller than the value assumed in this paper, a reduction of final $^7$Li abundance can be possible without spoiling all successes in standard cosmological models about elemental abundances and baryon-to-photon ratio.  This scenario requires injections of energetic photons with energies between the photodisintegration energy thresholds of $^7$Be (1.59 MeV) and D (2.22 MeV).  Although the fine tuning of the energy is necessary, it is a very simple solution to the $^7$Li problem.


Figure \ref{fig2} shows the constraints on this model in the ($\tau_X$, $m_X$) plane.  Contours correspond to
the adopted observational constraints on abundances.  The right region of solid lines for D, $^3$He, $^4$He,
and dashed line for $^7$Li are excluded by abnormal nuclear abundances.  The region bounded by two dashed
lines for D is excluded due to underproduction of D. $^7$Li is overproduced on the left of the solid 
$^7$Li line. The $^6$Li abundance is larger than that detected in the metal-poor star G020-024
in the right region from the curve for $^6$Li.  The right region from the dot-dashed line is excluded by large
entropy production. Note that, if the lifetime of the sterile neutrinos is shorter than about
$10^4$ s, then the decay occurs before the temperature drops to a value about $10^{-3}$ MeV and  
the  cutoff energy $E_C$ of the nonthermal photon spectrum mentioned just below Eq. (\ref{eq_spe}) will be about $2$ MeV. 
If $ 1.59 \> {\rm MeV} \le E_{\gamma 0} \le 2.22 \> {\rm MeV}$ the photon energy will be sufficient to break up $^7$Be into $^3$He and 
$^4$He, but deuteron will remain intact. As we mentioned above, if one can appropriately adjust the sterile neutrino number density, instead of using Eq. (\ref{5}), this would provide a solution to the $^7$Li problem. 


\begin{figure}
\begin{center}
\includegraphics[width=8.0cm,clip]{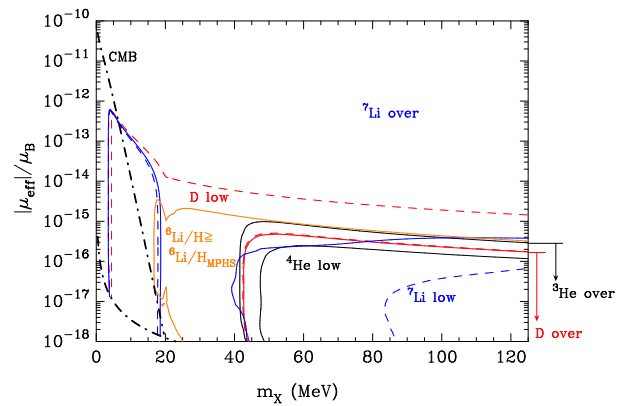}
\caption{The same constraints as in Fig. \ref{fig2} but for the ($m_X$, $|\mu_{\rm eff}|/\mu_{\rm B}$) plane.  The constraint from the baryon-to-photon ratio of WMAP is viable above the dot-dashed line located at the left bottom.  This line corresponds to the time of cosmological recombination.  \label{fig3}}
\end{center}
\end{figure}

\begin{figure}
\begin{center}
\includegraphics[width=8.0cm,clip]{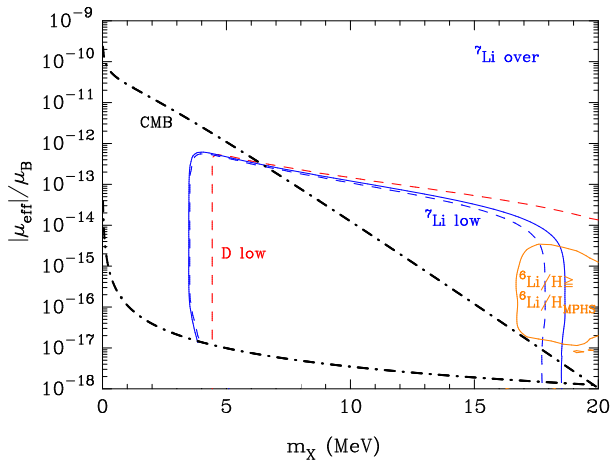}
\caption{The same constraints as in Fig. \ref{fig3} in a narrower ($m_X$, $|\mu_{\rm eff}|/\mu_{\rm B}$) plane.  \label{fig4}}
\end{center}
\end{figure}

Figure \ref{fig3} shows the constraints in the ($m_X$, $|\mu_{\rm eff}|/\mu_{\rm B}$) plane. Contours correspond
to the same observational constraints as in Fig. \ref{fig2}. The mass region between $0$ and $20$
MeV is enlarged in Figure \ref{fig4} for clarity. Higher values of the magnetic
moment correspond to shorter lifetimes whereas smaller values of the magnetic moment
correspond to longer lifetimes for the sterile neutrino [see Eq. (\ref{1})]. 
Since the sterile neutrinos we consider are
non-relativistic, their energy density decreases as $a^{-3}$ as the universe expands, where $a$ is the  scale factor of the universe. Photon energy density, on the other hand, decreases as $a^{-4}$. Hence right after the $e^+e^-$ annihilation epoch 
the photon density is larger than the sterile neutrino density and if
the sterile neutrinos decay too early during the expansion, the nonthermal photon density that they produce
will be a negligible fraction of the thermal photon density in the background and therefore will have no noticeable effect on BBN yields. 
This region corresponds to the upper part of the dash-dotted line which is nearly diagonal across Fig.
\ref{fig4}. The values of the magnetic moment above this line cannot be constrained from BBN considerations alone.
Below this line the sterile neutrinos live long enough for the
background radiation density to drop so that the nonthermal photons that they produce at sterile neutrino decay becomes increasingly significant. In this region, the magnetic moment can be constrained from BBN considerations. However, if the magnetic
moment is too low, i.e., the lifetime of the sterile neutrino is too long, then the decay happens after the
recombination epoch, i.e., $1.55\times 10^{13}$ s at $z=1088$ \cite{Hinshaw:2012fq}.  This is represented by the other dash-dotted line near the bottom of the Fig. \ref{fig4}.  The radiative decay occurring
after the recombination is directly observable today as a diffuse non-thermal background and is strongly constrained from measurements of $\gamma$-ray background and high energy
neutrinos \cite{1992NuPhB.373..399E}.

The low mass region of $m_X\lesssim 6.5$ MeV is constrained from WMAP measurement
of baryon-to-photon ratio, while the high mass region of $m_X\gtrsim 6.5$ MeV is constrained from
observational nuclear abundances.  

\begin{figure}
\begin{center}
\includegraphics[width=8.0cm,clip]{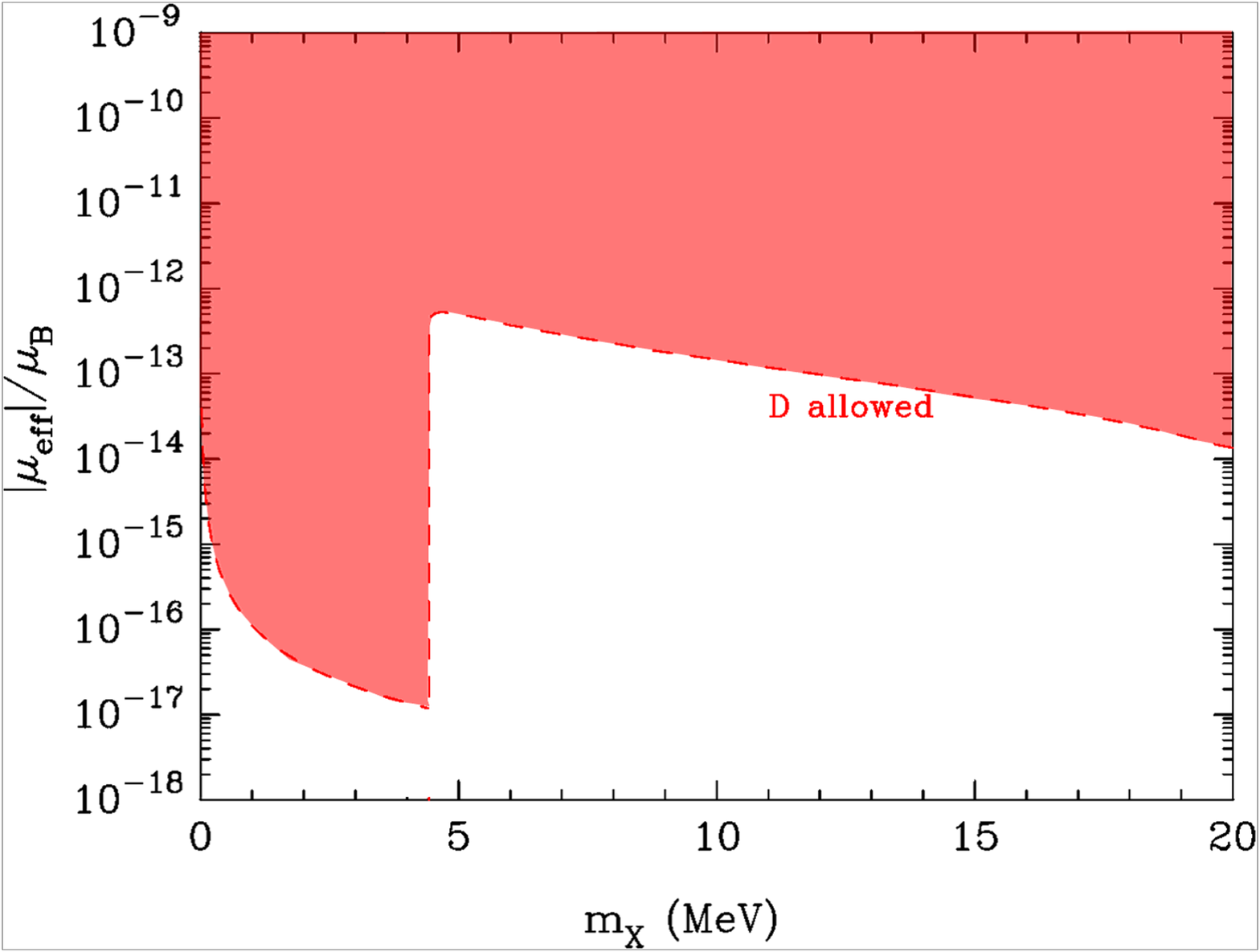}
\caption{Allowed region due to the constraint on D abundance in the ($m_X$, $|\mu_{\rm eff}|/\mu_{\rm B}$) plane.  \label{fig5}}
\end{center}
\end{figure}

\begin{figure}
\begin{center}
\includegraphics[width=8.0cm,clip]{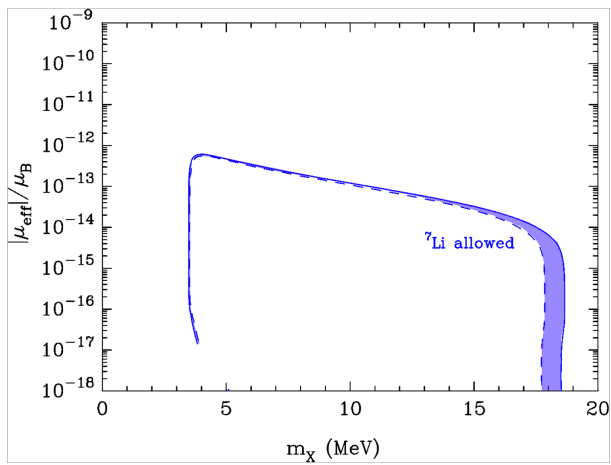}
\caption{Allowed region due to the constraint on $^7$Li abundance in the ($m_X$, $|\mu_{\rm eff}|/\mu_{\rm B}$) plane.  \label{fig6}}
\end{center}
\end{figure}

Figures \ref{fig5} and \ref{fig6} show allowed regions due to the constraints on abundances of D and $^7$Li, respectively.


In this paper we showed that observed light element abundances from the Big Bang can put significant constraints on the sterile neutrino masses and their contributions to the neutrino magnetic moment. 
A persistent puzzle of the modern cosmology is that there is no allowed region in which $^7$Li abundance is simultaneously consistent with observed abundances in metal-poor stars, the deduced value of the baryon-to-photon ratio from the CMB measurements and the BBN calculations. We showed that allowing a heavy sterile neutrino decay into active ones does not alter this situation. 
If one were to relax the requirement of using the CMB value of baryon-to-photon ratio or using the relic neutrino abundance deduced from the thermal freezeout of weak reaction we find a narrow-band region, determined from an overlap of allowed regions from D and $^7$Li  abundance, located  at $m_X \sim (3-5)$ MeV and $|\mu_{\rm eff}| \sim (10^{-17}-10^{-12}) \mu_B$. In this region, all light element abundances are consistent with the observed abundances, differently from abundances predicted in standard  BBN model. Of course such a sterile neutrino could also decay into electron-positron pairs, possibly further altering the nucleosynthesis yields. However, since we did not specify the interactions of this sterile object, it may be possible to suppress such decays in certain models. 
Note that if the sterile neutrino with $m_X \sim (3-5)$ MeV and $|\mu_{\rm eff}| \sim (10^{-17}-10^{-12}) \mu_B$ were viable it could place a limit on the electron neutrino magnetic moment. 
Taking that the electric dipole moment to be zero, assuming $\sum_i U_{ei} \sim 1$, where the sum is over all the active flavors, and noting that $U_{e4} <1$, this gives the limit of $\mu_{\nu} < 10^{-12} \mu_B$ for the diagonal magnetic moment of the electron neutrino which mixes with this sterile state. Of course it is very likely that $U_{e4}$ could be much smaller and the contribution of this sterile state to the neutrino magnetic moment is much less than the stated upper limit. 

\begin{acknowledgments}
ABB, YP, and MK would like to thank the NAOJ theory group for its hospitality during their visit. 
This work was supported in part by the NAOJ  Visiting Fellow Program (ABB, YP, and MK), in part by Grants-in-Aid for Scientific Research of the JSPS (200244035) and for Scientific Research on Innovative Area of MEXT (20105004), in part by the U.S. National Science Foundation Grant No. PHY-1205024, in part through JUSTIPEN (Japan-U.S. Theory Institute for Physics with Exotic Nuclei) under grant number DEFG02-06ER41407 (through the University of Tennessee), 
in part by the Scientific and Technological Research Council of Turkey (TUBITAK) under grant number 112T952, and 
in part by the University of Wisconsin Research Committee with funds
granted by the Wisconsin Alumni Research Foundation.
\end{acknowledgments}


%

\end{document}